\documentclass[final,1p,times]{elsarticle}
\usepackage{graphicx}
\usepackage{amssymb}
\usepackage{amsthm}
\usepackage{lineno}

\newcommand{\Kzs}{\mbox{$\mathrm {K^0_S}$}}

\journal{Nuclear Physics A}

\begin{document}

\begin{frontmatter}

  % Your Title - please insert
  \title{Meson and baryon femtoscopy in heavy-ion collisions at ALICE}

  %% Single author (and collaboration) - please insert
  \author{Maciej Pawe$\l$ Szyma\'nski (for the ALICE\fnref{col1} Collaboration)}
  \fntext[col1] {A list of members of the ALICE Collaboration and acknowledgements can be found at the end of this issue.}
  \address{Faculty of Physics, Warsaw University of Technology, Koszykowa 75, 00-662 Warsaw, Poland}

  %% Multiple authors
  % \author[auth2]{Marcus Junius Brutus}
  % \address[auth1]{Somewhere, Rome}
  % \address[auth2]{Somewhere else, Rome}

  \begin{abstract}
    We present results of femtoscopic analysis with $\Kzs\Kzs$, K$^{\pm}$K$^{\pm}$, pp, $\bar{\mathrm{p}}\bar{\mathrm{p}}$, $\mathrm{p}\bar{\mathrm{p}}$ and $\Lambda\bar{\Lambda}$ in Pb--Pb collisions at $\sqrt{s_{\mathrm {NN}}}=2.76$~TeV and $\Kzs\Kzs$ in pp collisions at $\sqrt{s}=7$~TeV registered by ALICE at the LHC. The femtoscopic radius is extracted from the one-dimensional correlation functions. The analysis reveals that the emission source sizes of pions, kaons and protons measured in heavy-ion collisions exhibit approximate transverse mass scaling which is consistent with hydrodynamic collectivity prediction. Furthermore, baryon-antibaryon correlations are investigated to study the influence of Final State Interactions (annihilation). It may contribute to lower multiplicities of protons observed at LHC energies, with respect to predictions from thermal models.
  \end{abstract}

\end{frontmatter} % do not change

%% linenumbers are useful for reviewing process
% \linenumbers

\section{Introduction}
The method of two-particle correlations at low relative momenta (commonly referred to as \emph{femtoscopy}) allows to extract space-time characteristics of the particle emitting source, created in heavy-ion collisions~\cite{kopylov1},~\cite{kopylov2}. Correlations of identical pions are usually used to perform this study. In particular, the ALICE Collaboration recently carried out the analysis of two-pion correlations in pp and central Pb--Pb collisions~\cite{alicefemtopp},~\cite{alicefemtoPbPb}.

The main goal of kaon and proton femtoscopic study is to complement information about the source size deduced from pion correlations. The analysis provides an extension of the transverse momentum range. Therefore, it is possible to verify whether the ``homogeneity lengths'' mechanism~\cite{sinyukov} is valid for mesons as well as baryons. Furthermore, one can test the transverse mass scaling of the source size which is interpreted as a signature of collective behaviour~\cite{lisa}.

Baryon-antibaryon correlations might be also used to study Final State Interactions (FSI) which can influence particle yields. It is observed that thermal models are not able to reproduce the p/$\pi^+$ ratio simultaneously with other hadron production ratios~\cite{preghenella},~\cite{milano}. It is argued that rescattering in the hadronic phase should be taken into account while determining particle yields~\cite{werner},~\cite{karpenko},~\cite{aichelin}. If the assumption is valid, it ought to  be reflected as anticorrelation in baryon-antibaryon correlation functions, due to the annihilation channel.

\section{Data analysis}
The dataset of Pb--Pb collisions at $\sqrt{s_{\rm NN}}=2.76$~TeV and pp collisions at $\sqrt{s}=7$~TeV consist of roughly 30 and 300  million events, respectively. Tracks were reconstructed with the Time Projection Chamber (TPC)~\cite{alicesummary}. Particle identification of K$^{\pm}$, p and $\bar{\mathrm{p}}$ was performed with the TPC and the Time-of-Flight detector~\cite{alicesummary} in $|\eta|~<~0{.}8$. Primary tracks were selected based on the Distance of Closest Approach to the primary vertex (DCA). $\Lambda$ ($\bar{\Lambda}$) candidates were extracted by identifying daughter tracks using TPC and TOF detectors. The decay channel $\Kzs~\rightarrow~\pi^+\pi^-$ was used for identification of neutral kaons. More details about $\Kzs\Kzs$ analysis can be found in~\cite{kzspaper}.

Particles from the same event were paired to form the signal distribution of relative momenta $q_{\mathrm{inv}}=2 \cdot k^{*} = \sqrt{(p_1-p_2)^2-(E_1-E_2)^2}$, from two different events to form a mixed background distribution. Pair cuts based on the ratio of clusters (groups of detector signals) shared by two tracks to all clusters of both tracks and the angular distance inside TPC between two tracks were used to account for fake low-q pairs (splitting) and two-track inefficiency (merging). After pair selection, the resulting signal distribution was divided by the background distribution to form the correlation function.

\section{Kaon femtoscopy}
$\Kzs\Kzs$ correlations were fitted with a parametrization which includes Bose-Einstein statistics as well as strong FSI~\cite{lednicky}. Femtoscopic radii from charged kaon correlations were obtained using the Bowler-Sinyukov formula~\cite{alicefemtopp}.

\begin{figure}[h]
        \centering
        \includegraphics[width=0.47\textwidth]{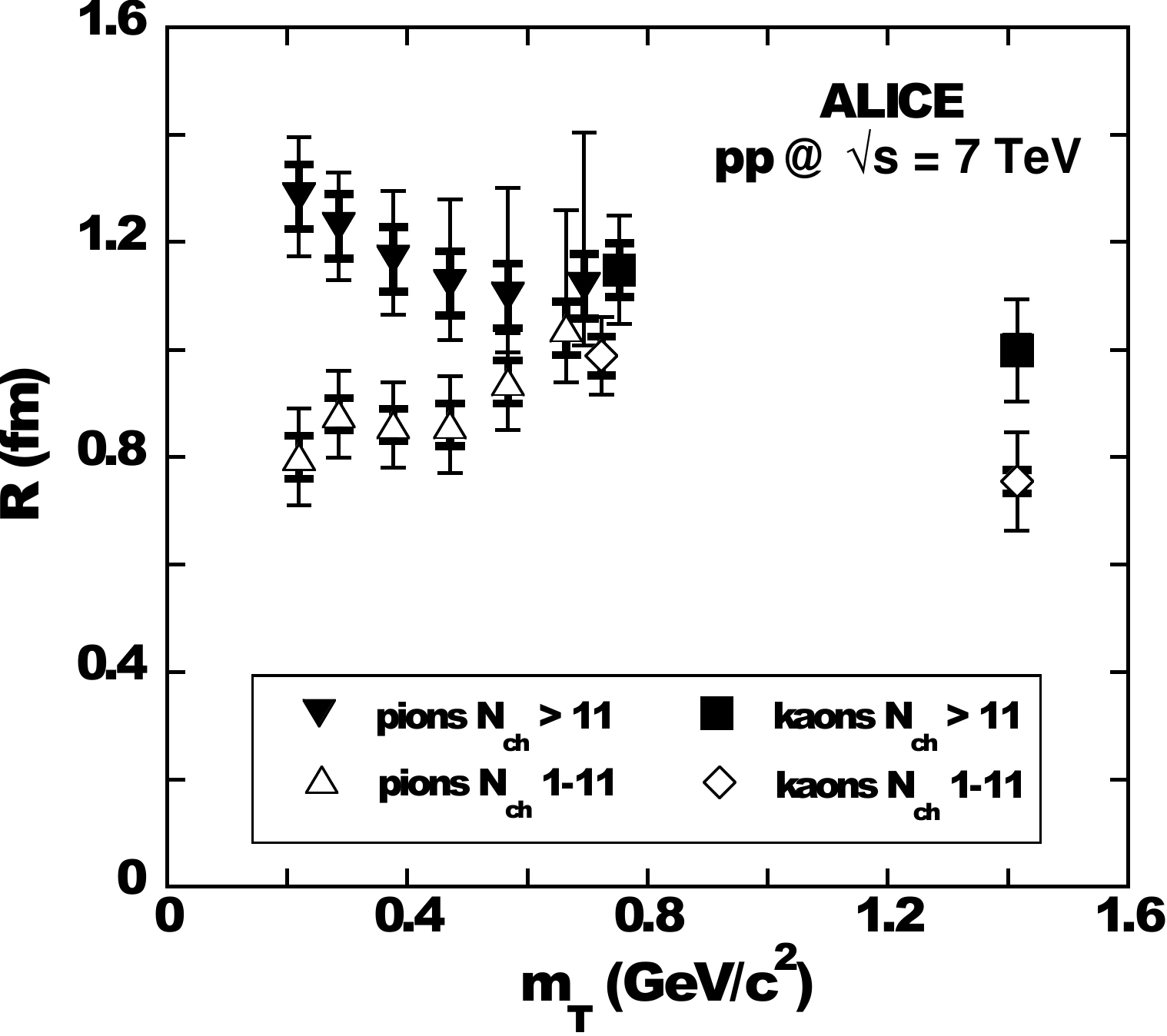}
        \includegraphics[width=0.52\textwidth]{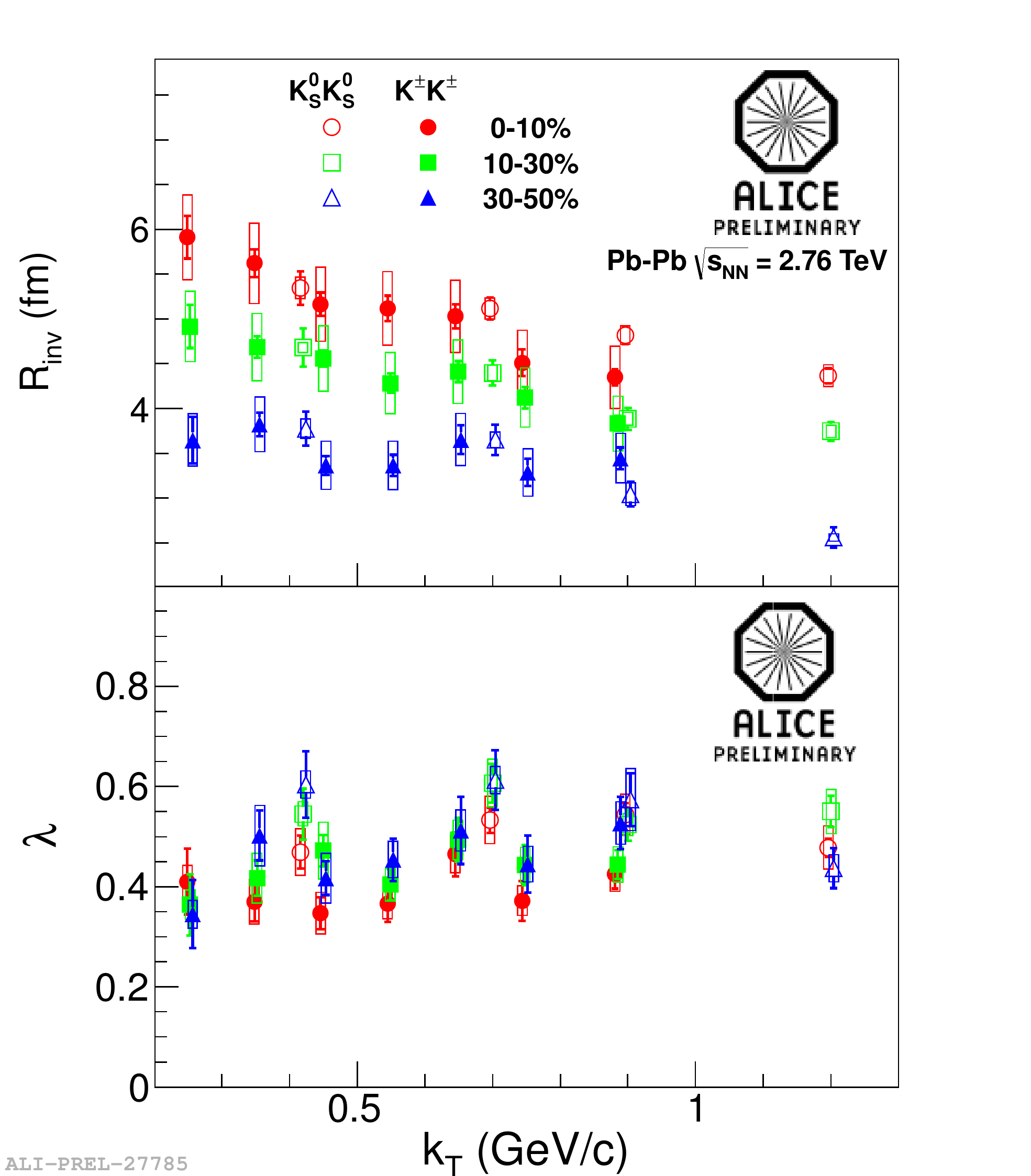}
         \caption{ \footnotesize  Left panel : Transverse mass dependence of the radii extracted from neutral kaon and pion femtoscopy in pp collisions at $\sqrt{s}=7$~TeV~
\cite{kzspaper}. Right panel: the comparison of $R_{\mathrm {inv}}$ and $\lambda$ vs. $k_{\mathrm T}$ for neutral and charged kaons correlations in Pb--Pb collisions at $\sqrt{s_{\rm {NN}}}=2{.}76$~TeV.}
        \label{kzspp}
      \end{figure}

The transverse mass dependence of the radii extracted from $\Kzs\Kzs$ and $\pi^{\pm}\pi^{\pm}$ correlations in pp collisions (presented in the left panel of Fig.~\ref{kzspp}) appears to be weak. Also, the radii increase with increasing event multiplicity.

The comparison of the radii obtained from K$^{\pm}$K$^{\pm}$ and $\Kzs\Kzs$ femtoscopy in Pb--Pb collisions is shown in the right panel of Fig.~\ref{kzspp}. The results from the two analyses are consistent with each other. The radii clearly increase with the event multiplicity and decrease with pair transverse momentum ($k_{\mathrm T}=|\vec{p}_{\mathrm{T,1}}+\vec{p}_{\mathrm{T,2}}|/2$) which is compatible with the hydrodynamic description. The value of the correlation strength $\lambda$ is of the order of $\sim 0.5$ and it does not show any $k_{\mathrm T}$ dependence.

\section{Baryon femtoscopy}
\begin{figure}[h]
        \centering
        \includegraphics[width=0.495\textwidth]{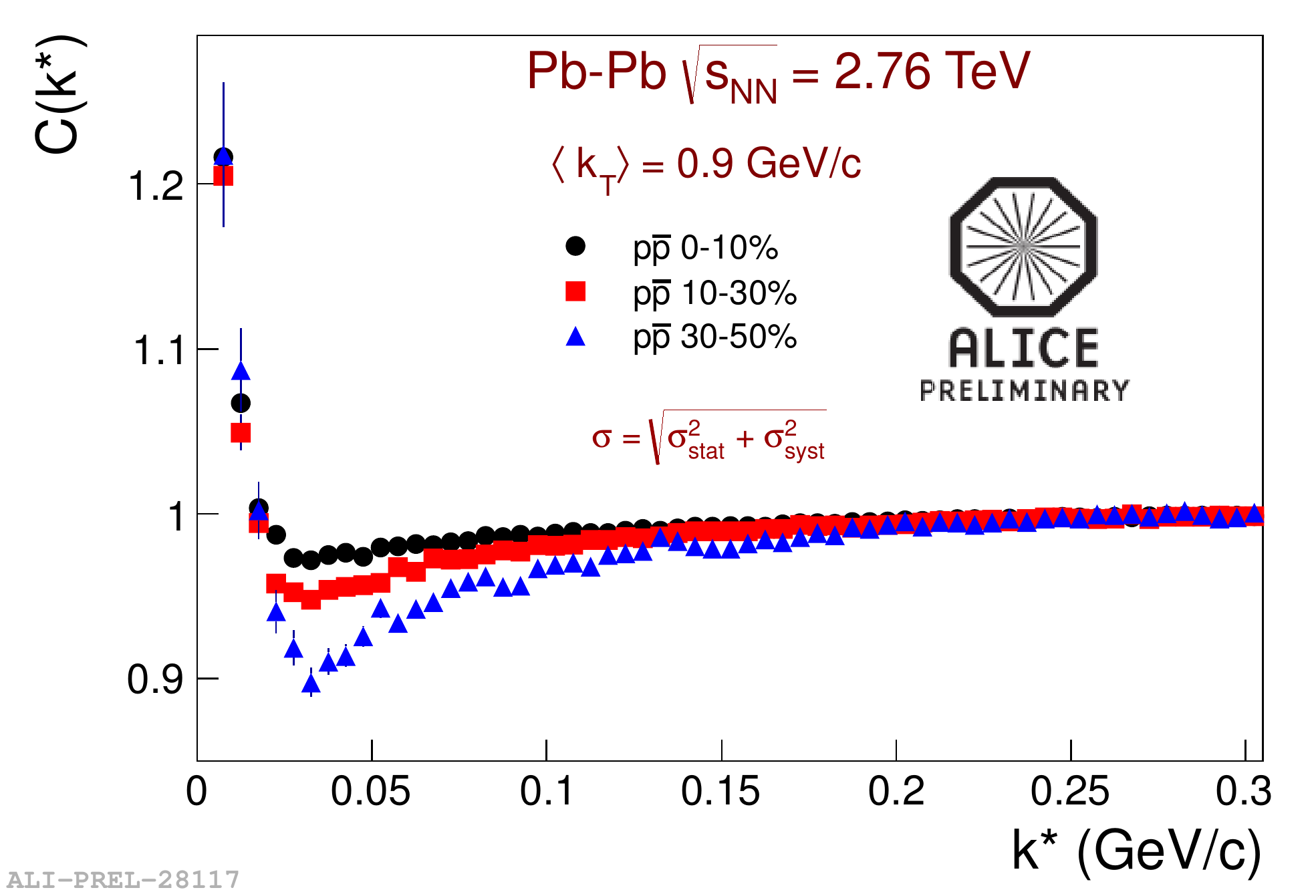}
        \includegraphics[width=0.495\textwidth]{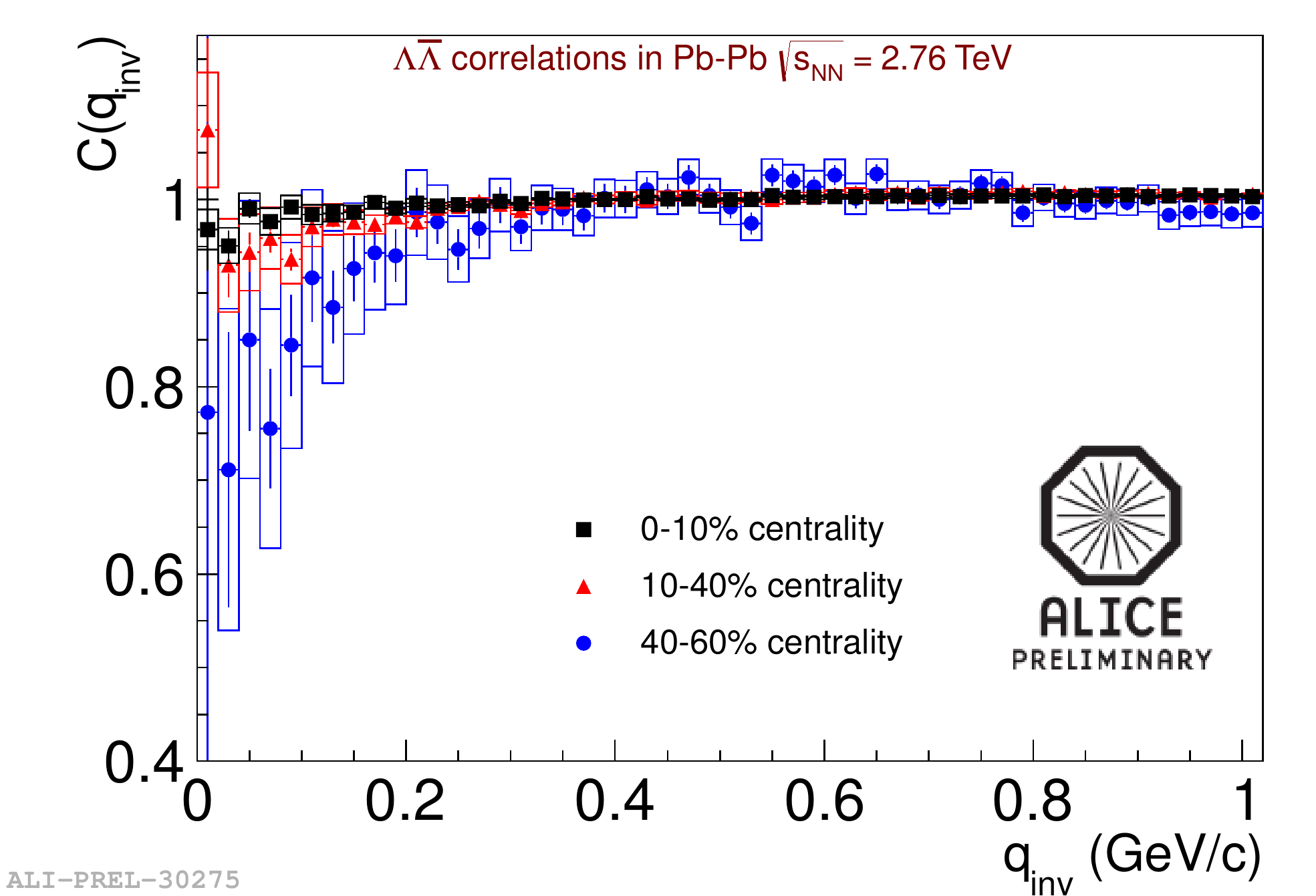}
        \caption{ \footnotesize  $\mathrm{p}\bar{\mathrm{p}}$ (left panel) and $\Lambda\bar{\Lambda}$ (right panel)  correlation functions from the Pb--Pb collisions at $\sqrt{s_{\mathrm {NN}}}=2.76$~TeV.}
        \label{cfBAB}
      \end{figure}

The p$\bar{\mathrm{p}}$ correlations (left panel of Fig.~\ref{cfBAB}) show a maximum for the low relative momentum due to Coulomb attraction, then a wide minimum caused by the annihilation part of the strong FSI. % The femtoscopic effect is wide comparing to $\mathrm{p}\mathrm{p}$, providing a better statistical handle on the system size.
$\Lambda\bar{\Lambda}$  correlation functions (right panel of Fig.~\ref{cfBAB}) show a clear suppression at low relative momentum for three centralities ranges. It is consistent with annihilation in the strong FSI between baryons and antibaryons. The strength of the correlation effect increases with more peripheral events, an indication that the emitting source size is shrinking for those events.

The correlations of pp and $\bar{\mathrm{p}}\bar{\mathrm{p}}$ pairs (left panel of Fig.~\ref{mtscaling}) are due to a combination of Fermi-Dirac statistics, Coulomb and Strong FSI which results in a distinct maximum for $q_{\mathrm {inv}}~\approx~40$~MeV/$c$ \cite{lednicky}. The height of this peak is related to the femtoscopic radius.

Due to the fact that feed-down from weak decays cannot be neglected in high-energy heavy-ion collisions, the residual correlations from p$\Lambda$ system in pp correlations have to be taken into account. Since  $\Lambda$ decays into p and $\pi^{-}$ with small decay momentum with respect to the mass of p, femtoscopic correlations between a primary p and a $\Lambda$ might still be detected for a pair consisting of the primary p and the p from $\Lambda$ decay. The method of simultaneous fitting of $\mathrm{p}\mathrm{p}$ ($\bar{\mathrm{p}}\bar{\mathrm{p}}$, $\mathrm{p}\bar{\mathrm{p}}$) and $\mathrm{p}\Lambda$ ($\bar{\mathrm{p}}\bar{\Lambda}$, $\mathrm{p}\bar{\Lambda}$) correlations was used (left panel of Fig.~\ref{mtscaling}).

\section{Transverse mass dependence of the radii}
\begin{figure}[h]
        \centering
        \includegraphics[width=0.49\textwidth]{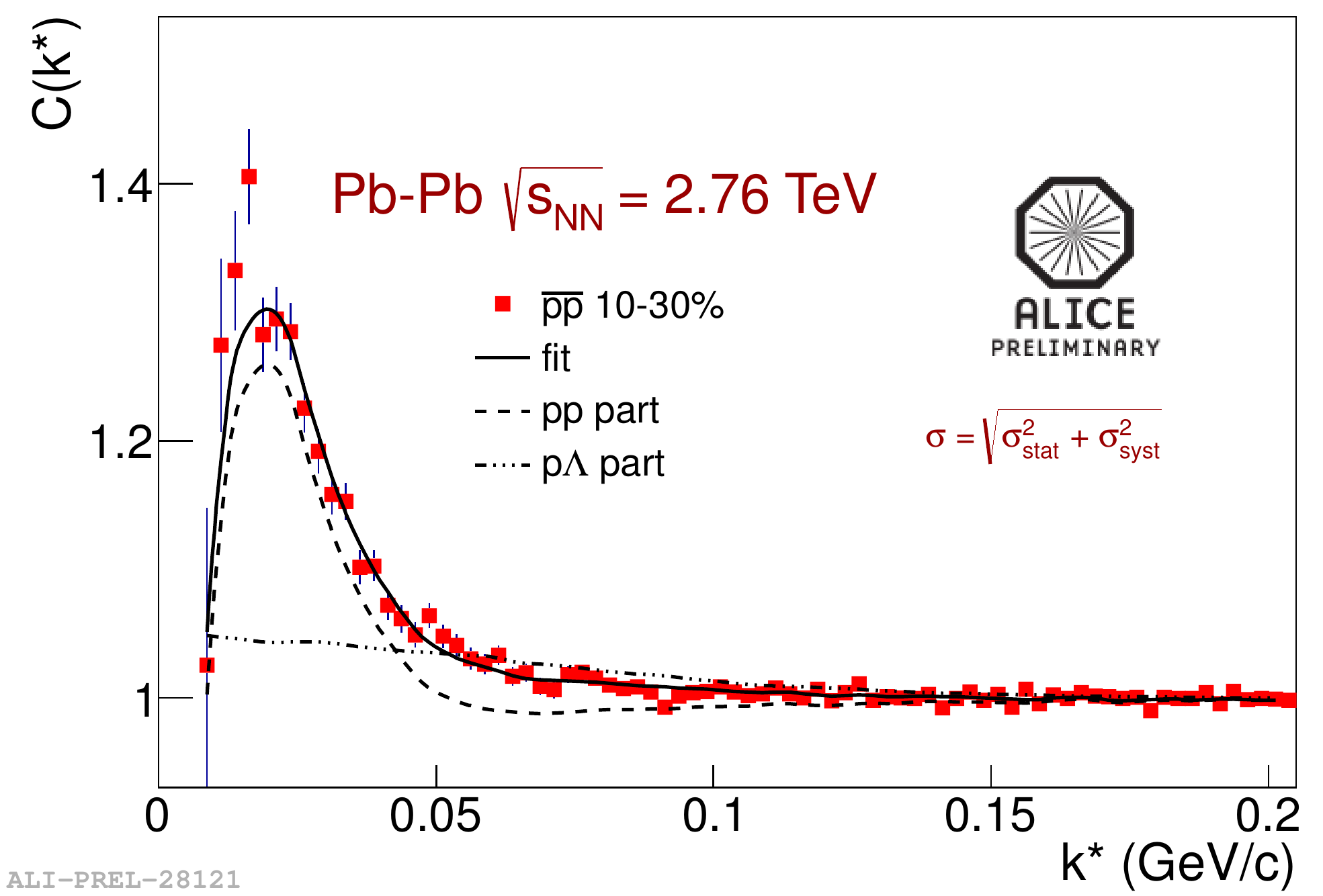}
        \includegraphics[width=0.50\textwidth]{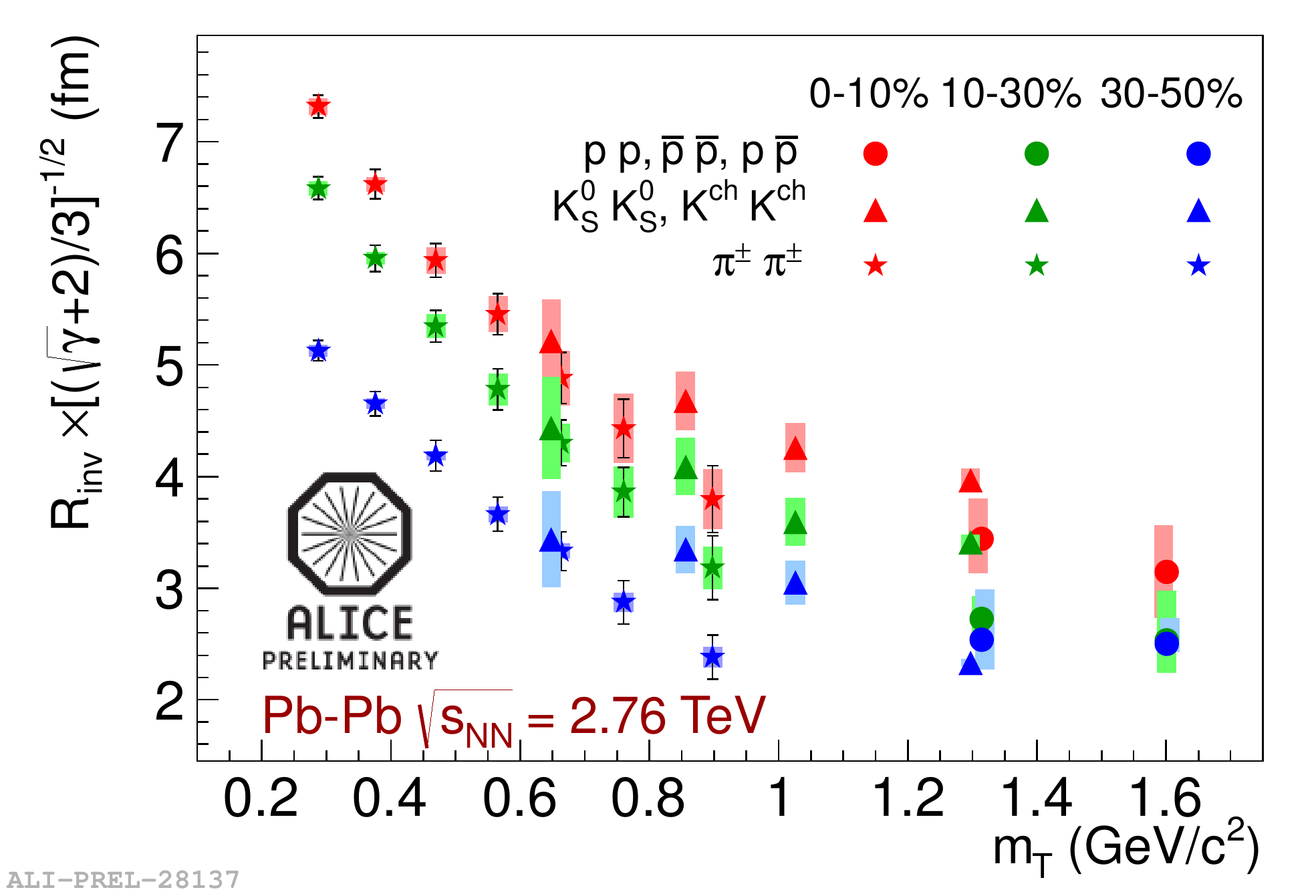}
        \caption{ \footnotesize Left panel: example of the fit to the $\bar{\mathrm{p}}\bar{\mathrm{p}}$  correlation functions, taking into account contribution from residual correlations (see text for details). Right panel: $m_{\rm T}$ dependence of the radius parameter extracted from correlations of pions, charged kaons, neutral kaons and protons.}
        \label{mtscaling}
      \end{figure}

Transverse mass scaling of the femtoscopic radii is considered as a signature of collective behaviour of the system created in heavy-ion collisions. The right panel of Fig.~\ref{mtscaling} shows transverse mass dependence of the invariant radii scaled by kinematic factor obtained from $\pi$, K and p femtoscopy. As shown, the radii seem to follow the same curve which is in agreement with hydrodynamic prediction.% However, more precise study is needed to 

\section{Summary}
The femtoscopic parameters for the radius of the emission source sizes of kaons and protons are extracted from one-dimensional correlation functions. Results  show an increase of the radius with increasing multiplicity and slight decrease of the radius with increasing  pair transverse momentum.

The observed significant annihilation in baryon-antibaryon correlations may be responsible for the decrease of proton yields at LHC energies with respect to thermal models.

Approximate $m_{\rm T}$ scaling of the femtoscopic radii for pions, kaons and protons is observed.% Such a tendency enhances the picture of the strongly-interacting system created in heavy-ion collision.

% \section*{References}

\end{document}